# A Navigational Approach to Health

**Ramesh Jain**
**Department of Computer Science**
**University of California**
**Irvine, CA 92697**

jain@ics.uci.edu

**Abstract**

What if an app could guide you to better health, similar to how GPS navigation directs you to your desired destination? What if the app could use real-time information to redirect you around a disease, just as you're rerouted to avoid traffic? What if the app could provide step-by-step directions to get you to your optimal health state, whether you're a professional athlete or retired school teacher? We discuss how this navigational approach to healthcare could become a reality by combining emerging technology with well-established cybernetic principles.

## Introduction

A novel approach to caring for human health is in the works, resulting from the metanexus of new developments in genetics, biology, sensing, big data and smartphones. The goal is to use artificial intelligence — including machine learning, augmented reality and gamification, diverse sensors, mobile phones, and data management — to implement a *navigational approach* that builds on the strengths of cybernetics. This approach should disrupt the healthcare industry by providing more effective and efficient personal health management. It will facilitate the application of medical knowledge to guide people in adapting their lifestyle, environment and socio-economic situation to better maintain their health and maximize their quality of life.

## Building on Cybernetics

Cybernetic principles [1], proposed by Norbert Wiener as a mechanism for control and communication in machines as well as living systems, have transformed the design of complex systems. Continuous measurements are a key component in closed-loop feedback control, which is essential for implementing systems that work in real-world noisy environments. From a simple thermostat-based air-conditioning system in a home, to soon-to-be-common autonomous cars, these principles are omnipresent.

In the simplest terms, cybernetics is about setting goals and devising action sequences to accomplish and maintain those goals in the presence of noise and disturbances (see Figure 1). Initially proposed as guiding principles for machines and living systems, Wiener later suggested that many of these principles are applicable to social systems as well [2], and in a recent book [3], John Markoff traces many breakthroughs in AI to cybernetics. Some of the currently popular topics in AI, such as machine learning, adaptive systems, and robotics, were first mentioned in cybernetics.

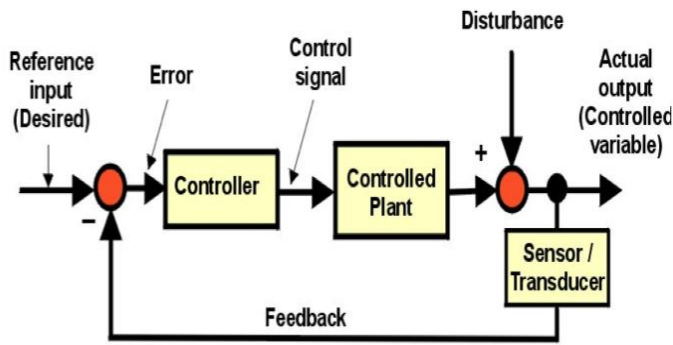

**Figure 1:** The main component of cybernetic systems is feedback for controlling the real-time behavior of complex real-world systems. In these systems, the destination (or desired state) is perpetually compared with the current state, and the discrepancy or error is used to guide the system.

**Exploring the Transformative Effects of Cybernetization**

Wiener saw the transformative power of his invention early on. His concerns about the social implications of automation in industry and the possibility of machines becoming more powerful than humans led him to stop pursuing further research in this area [3]. However, applying cybernetization to otherwise open systems presents many opportunities to develop humanitarian systems and other systems that could improve people's quality of life.

The cybernetization of an existing approach is enabled by the availability of sensors that can estimate the system state and perpetually feed this information back to the system, letting it generate new control signals as required to move toward the desired goal or destination. Navigation systems present a good example of the effects of cybernetics.

Such systems eventually replaced maps, which date all the way back to about 2300 B.C. Over the years, maps evolved with technology [4], and at the start of the 21$^{st}$ century, online maps became very popular. Powerful algorithms let people plan their routes using these digital maps, so users could print turn-by-turn directions for optimal routes based on different requests, such as "minimize the distance traveled" or "avoid highways and toll-roads" (see Figure 2).

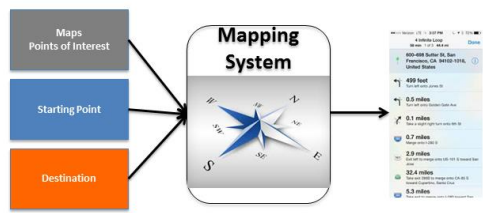

**Figure 2:** Computerizing maps made it easy to use algorithms to find the optimal route between two points and to generate turn-by-turn directions.

The real revolution, however, came with the prevalence of sensing technology. When GPS became less expensive, more readily available and sufficiently accurate, it started appearing in not only cars but also mobile phones and other devices, offering current location data. The maps and navigation systems improved rapidly between the early 1990s and mid 2000s [5], as higher quality maps and interfaces became available. Once crowd sourcing and other approaches grew powerful enough to determine real-time traffic, the navigation space was "cybernetized," as shown in Figure 3. Navigation systems then became so effective and easy to use that most people stopped using actual maps and started expecting their mobile phones or car dashboards to present step-by-step audio guidance, with automatic rerouting based on real-time information. These sophisticated cyber-physical-human systems apply geo-spatial and temporal "knowledge captures" in maps to continuously update vehicle models using GPS and other sensors, and they present the data using user-friendly computer-human interfaces [6].

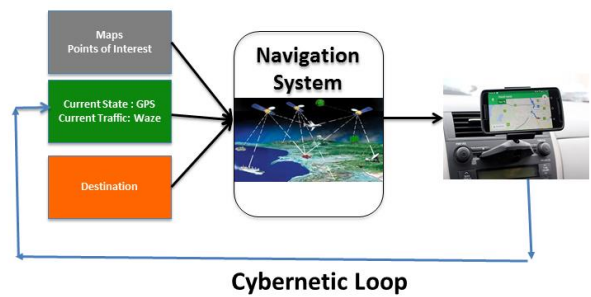

**Figure 3:** The cybernetization of navigation systems required automatically knowing the current location and using it perpetually to determine routes, taking traffic and other factors into consideration.

This is the power of cybernetics — it lets a system remain effective even in presence of noise and varying conditions.

**Applying Cybernetics-Directed Navigation to Healthcare**

The human body and other biological systems have an intricate play of real-time sensors and actuator functions. Homeostasis is based on similar feedback principles. When something within the human body fails to function properly, the unstable condition can result in a disease, and we try to cure this condition by fixing the defective components in the communication/control loop.

In current medical practice, the process of fixing a disease state and restoring the normal health state is based on alerts from the person. However, in some cases, these alerts are too late, resulting in serious consequences. In other cases, the alerts remain below the level of consciousness for the person, but they are usually manifested in some biological markers that can be measured using sensors. These measurements can be used to estimate the person's health state, and, if required, corrective actions can be taken to return the person to the desired state. Such actions might include making lifestyle or environmental changes or taking medication. These actions would be determined based on biological and medical knowledge and a model of the person.

This sensing loop for perpetually measuring the person's health state and taking corrective action brings cybernetics to health management. Predictive, personalized precise and preventive medicine aims to identify chronic diseases that remain prodromal [7] for some time. Applying techniques to detect and address a disease in its early stages [8] can help delay or prevent the disease.

Progress in technology can extend the cybernetic loop outside a person's body using emerging sensors and following cybernetic principles of continuous monitoring for proper communication and control. AI is very useful in enabling this loop, and machine-learning-based tools facilitate diagnosis and help estimate health states. Contextual reasoning will also help by guiding the person to take corrective actions based on all available medical and environmental knowledge.

**Disrupting the Symptom-Driven Health Cycle**

Currently, most people think of their health only when they are sick. When they feel under the weather, they visit a doctor. A successful doctor's visit results in the diagnosis of the person's health state (possibly a disease) and a prescription for medicine and any other regimen that will help cure the disease or at least ease the discomfort. Usually, the doctor's office then schedules a follow-up visit.

Consider this process from the patient perspective:

1) You are not feeling good. Some of your health state parameters are not in their normal range, resulting in an uneasy and abnormal state. Once the state is beyond your tolerance, you reluctantly see a doctor.
2) An assistant takes routine measurements of biomarkers, including weight, temperature, heart rate, and blood pressure. When the doctor comes, she usually listens to signals inside your body using a stethoscope and asks questions related to your health status and your family and health history. She is trying to estimate your current health state and build and update your personal model.
3) To get a better understanding of your health parameters, the doctor further examines your body and checks your vital signs. If these measurements don't provide sufficient data, the doctor orders imaging or pathology tests to obtain more information about your internals.
4) Combining all observations and using all medical knowledge, the doctor estimates your current health state, characterizes it as one of the known disease states and determines its level of severity.
5) The doctor uses your personal model and current health state, her own medical knowledge about diseases, and general environmental knowledge to recommend corrective actions in the form of prescriptions and a regimen. These may involve medication, lifestyle or environmental changes, or some other treatment.
6) You try to follow the regimen. Taking medications is easy; making lifestyle changes (such as controlling your diet) is often more difficult. There is generally no mechanism to verify compliance. The assumption is that you are interested and able to comply, but there are not many good approaches to remind people about compliance and adherence.
7) Depending on the severity of the disease, the doctor might want to see you periodically to repeat steps 2–6 as needed. The period depends on criticality: you might revisit the doctor's office, be admitted to the hospital or receive critical care.

What does this process reveal? The lifeblood of your health is your data.

Data is converted into actionable information. Knowledge from different fields, particularly from the medical sciences and about the environment, is used for estimation and recommendations. Data comes in a variety of forms and is measured using different "sensors," such as personal feelings, vital-sign measurements, imaging and pathology reports, the doctor's estimation based on the data collected via his or her audio-visual-tactile senses, and many more.

In this process, the first three steps involve data acquisition. Step 4 is state estimation, step 5 is recommendations from a professional using knowledge of the field, step 6 is compliance by the patient and step 7 is follow-up or repetition at regular intervals to manage further complications. These steps form a MEGI cycle: Measure, Estimate, Guide and Influence, as shown in Figure 4. This cycle has evolved over a long period and has stabilized to its current form as discussed here.

During the Measure step, data is collected from many relevant sources, and, using that data, the doctor estimates the patient's health state and diagnoses the problem.

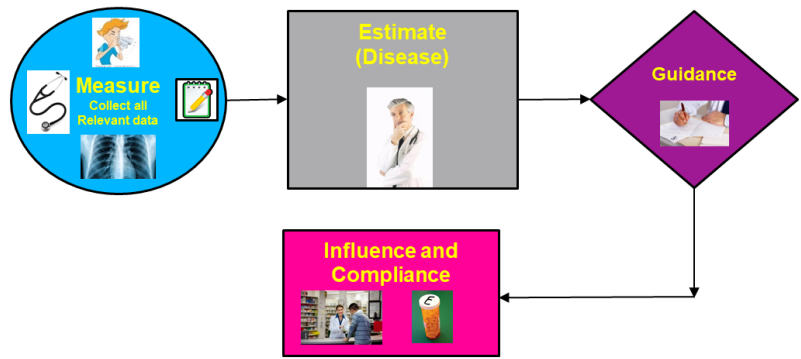

**Figure 4:** The Measure, Estimate, Guide, Influence (MEGI) cycle is the norm. It has many built-in delays and components that could work more effectively if AI and data science were used with the right sensors. Another problem is that this is an open-loop system.

Diagnosis is usually in the form of a disease and its level of severity. This Estimation step is what is most important in current healthcare. Once you get the right diagnosis, medical and other knowledge guides the prescriptions and regimens.

The Guide step is to personalize these prescriptions and regimens based on the specific patient and his or her personal situation.

Doctors encourage their patients to follow the prescriptions and regimens and usually request a follow-up visit, at which point the cycle is repeated. This can occur within months, weeks or days, or in severe cases, the person may be admitted to hospital for close and frequent repetition of the cycle. In extreme cases, a surgery followed by admission to an Intensive Care Unit may be advised.

**Cybernetizing the MEGI Cycle**

Can we "cybernetize" the MEGI cycle — that is, apply the principles of cybernetics to the cycle? In the current MEGI cycle, there are many human elements, and the ultimate operation is not a closed-loop system. Many measurements involve humans — the patient, doctor, pathologists and radiologists and so on. The guidance and influence parts also involve doctors and other people. So, can we apply a cybernetics approach to personal health to gain the benefits of self-regulation, as has happened in many other complex systems?

Given how progress in sensors, computing and AI has disrupted so many fields, it's natural to ask whether MEGI can be cybernetized. Personal healthcare is one of the most important areas for disruption because of the implication for our quality of life. The financial impact is also likely to be staggering.

Common sensors (accelerometers, gyroscopes, GPS, cameras and microphones) are now found in almost every smartphone, and other measurement technologies are appearing, including those that measure temperature, perspiration, heart rate, sleep, galvanic skin resistivity, blood oxygen, blood sugar and blood pressure, making cybernetic health possible.

The cybernetization of the MEGI cycle will result in a major disruption in personal health.

**Creating a Personal Health Navigator**

Let's now map the seven steps discussed above in a cybernetics version of the MEGI cycle with the following components.

**Measurement:** Using a smartphone and augmenting wearable sensors already on the market with emerging sensors that can measure many bodily functions, we could identify normal bodily parameters to determine the health state. A device, likely a smartphone or its equivalent, would continuously collect all of these measurements. This data would measure most of the information related to the person's health state. Only under very specific situations would special measurements be required. (Steps 1-3 above.)

**Estimation:** All those measurements could be used to estimate the person's health using mathematical as well as medical knowledge-based techniques. This estimation could indicate proximity to a disease. The goal of estimation would be to measure your health state without assigning the state to a specific disease or other semantic labels. Estimation techniques have been of great interest in the design of many applications [9]. (Step 4 above.)

**Guidance:** If the user wanted to change his or her health state, the user would issue a request and the system would use medical, environmental, and other relevant knowledge sources and the personal health state to provide the right guidance in terms of lifestyle or environmental changes or medications for getting to the desired state. This guidance would be perpetual until the user achieved the desired state. Guidance at each moment is very similar to contextual recommendation systems, which are currently of great interest to AI researchers [10]. (Step 5 above.)

**Influence and Adherence:** Providing guidance alone is not enough. Guidance must be situationally actionable and easy to follow. In most cases, mechanisms for nudging, incentivizing and inspiring might be required, along with subtle approaches for measuring compliance. (Steps 6-7 above.)

What is equally important is to implement this cycle on an almost continuous basis by performing all of these steps frequently to make sure that the person's health state remains in a safe zone. This implementation rate can be directly determined by the system.

The above steps are for dealing with non-emergency health situations. The system could function autonomously most of the time but could guide the user to a nearby resource in an emergency situation.

### Addressing the Challenges

Of course, numerous challenges must be addressed before we can reap significant benefits from cybernetic health. Here, I briefly discuss components that must be developed to implement a personal health navigation system (see Figure 5).

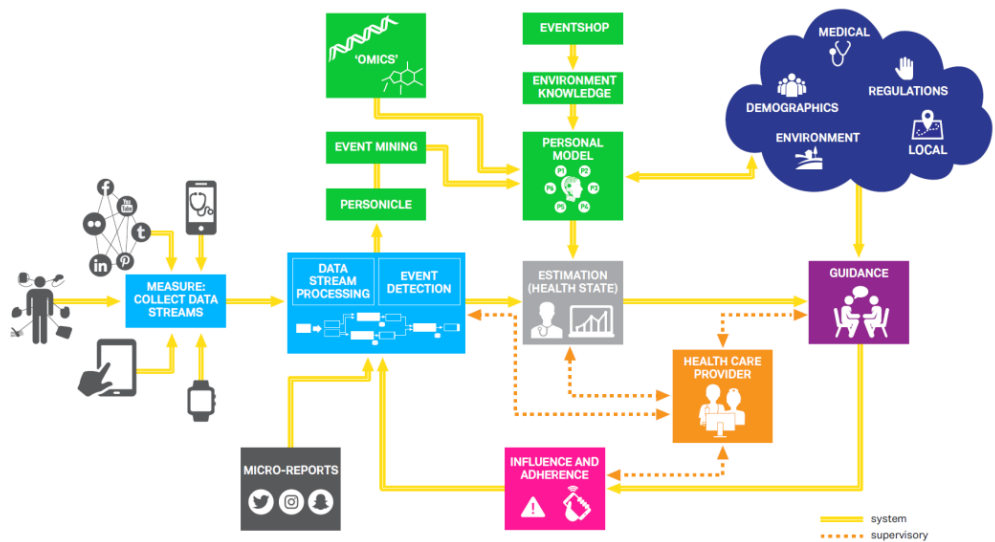

**Figure 5:** A health navigator is a closed-loop system based on continuous measurement. Guidance is based on a knowledge-based recommendation engine that uses the state of the system. Healthcare professionals may provide a supervisory role in an otherwise autonomous system.

**Personal Models:** Each individual is a unique system that must be modeled to effectively estimate the person's health state and provide precise guidance. The personal model captures how a person reacts to different stimuli under specific conditions.

To model a person, some relatively longer-term information comes from her genome modulated by the proteome, transcriptome and epigenome and reflected in the metabolome. Lifestyle, environment and socio-economic factors also play an important role in building a model of the person. This model is not static; it changes with age and other life conditions, so model building is a dynamic process.

**Health States Estimation:** Based on measurements of different biomarkers, the health state of a person must be estimated. In current medical practice, this step is the same as diagnosing a disease condition. To better understand health, more quantitative approaches require defining more health state objectives. The health state can be classified as a disease state in the same way as a combination of basic color components may be called pink or purple, but to manage colors more precisely, one must consider the primary components. A person's health state characterizes health objectively and can be assigned different symbolic labels, like diabetes. To implement predictive, preventive and precise medicine, it is important to objectively characterize health. Disease-centric estimation looks at the health state through colored glasses and is likely to result in biased decisions.

Estimation techniques for health states will require deep biological knowledge. Formal state-space models may emerge over time with associated observability and controllability conditions. In the interim, however, we might need to build rule-based modeling techniques to implement other aspects of the complete system.

**Situationally Actionable Knowledge:** Prescriptions and specifications of regimens to deal with diseases are common techniques currently used in medicine. These techniques are based on available medical knowledge and other relevant knowledge sources, including environmental knowledge. Effectively, all knowledge sources are organized to find and recommend appropriate actions in a given situation.

These knowledge sources may need to be organized at a finer granularity to deal with changing the health state rather than merely getting the person out of the disease state. Moreover, depending on the perspective of the designer, different knowledge sources may be used.

**Techniques to Encourage Compliance:** Once a recommendation about lifestyle and medications is made, the person is responsible for following up and complying with the specifications. As is well known, for various reasons, influencing people to follow the recommendations is challenging. Techniques must be developed to influence people and help them with compliance. [11].

**Exploiting Opportunities**

Disruptive transformations in health have become a possibility because of the metanexus of biology, genetics, sensors, computing, and mobile devices. Progress in these areas opens up the possibility of building personal models and using them in a cybernetic framework to develop personal health navigators. In addition to the obvious advantages of a closed-loop feedback system, the ability to measure biomarkers and take frequent — even instantaneous — actions allows this personal navigation system to function satisfactorily even when the model and measurements may not be perfect.

Another major opportunity is created by the smartphone, which is fast becoming a surrogate of the person. Smartphones could provide measurements and alerts and suggest certain actions.

Furthermore, all of the data collected for an individual could be shared and aggregated to build powerful population models related to diseases to provide recommendations in different situations.

Effectively, we have an opportunity to help people better manage their health while advancing our understanding of diseases and expediting research of potential cures. The availability of such massive amounts of individual and population-level data is unprecedented. We need to build on it, navigating society toward a healthier future.